\documentclass[11pt]{cernrep}
\usepackage{graphicx}
\usepackage{amsmath}
\usepackage{graphics}

\def\be{\begin{equation}}
\def\ee{\end{equation}}

\begin{document}
\title{Perturbative description of inclusive single hadron production at HERA}
\author{S.~Albino}
\institute{{II.} Institute for Theoretical Physics, University of Hamburg,\\
             Luruper Chaussee 149, 22761 Hamburg, Germany}
\maketitle
\begin{abstract}
Light charged hadron production data in the current fragmentation region at HERA
are calculated using next-to-leading order perturbative calculations and fragmentation functions
obtained from similar data from $e^+ e^-$ reactions.
General good agreement is found at large photon virtuality $Q^2$ and intermediate momentum fraction $x_p$, 
consistent with fragmentation function universality.
The description of the small $x_p$ and $Q^2$ region is improved by incorporating hadron mass effects.
\end{abstract}

\section{Introduction}

Unpolarized quark fragmentation functions (FFs) for charge-sign unidentified light charged hadrons $h^\pm=\pi^\pm$, 
$K^\pm$ and $p/\overline{p}$
have been well constrained by data from $e^+ e^- \rightarrow h^\pm+X$ using calculations at 
next-to-leading order (NLO) accuracy. 
Due to universality in the factorization theorem, such FFs can be used to calculate the similar measurements of
$ep\rightarrow e+h^\pm+X$.
This contribution summarizes the main results of \cite{Albino:2006wz} comparing $ep$ reaction
data in the current fragmentation region 
from the H1 \cite{Adloff:1997fr} and ZEUS \cite{Breitweg:1997ra}
collaborations at HERA with calculations using FFs extracted from $e^+ e^-$ reactions.

The kinematic degrees of freedom are chosen to be the centre-of-mass energy $\sqrt{s}$ of the
initial state $ep$ system,
the magnitude of the hard photon's virtuality
$Q^2=-q^2$, the Bjorken scaling variable $x=Q^2/(2P\cdot q)$ and the scaled
detected hadron momentum $x_p=2p_h \cdot q/q^2$. 
The normalized cross section (with the $s$ dependence omitted for brevity) takes the form
\be
F^{{\rm proton}\ h^\pm}({\rm cuts},x_{pA},x_{pB})
=\frac{\int_{\rm cuts} dQ^2 dx \int_{x_{pA}}^{x_{pB}} dx_p 
\frac{d \sigma^{{\rm proton}\ h^\pm}}{dx_p dx dQ^2}(x,x_p,Q^2)}
{\int_{\rm cuts} dQ^2 dx \frac{d \sigma^{\rm proton}}{dx dQ^2 } (x,Q^2)},
\label{XSmeas}
\ee
where ``cuts'' refers to a specified region in the $(x,Q^2)$ plane, and where $x_{pA(B)}$
is the lower (upper) edge of the $x_p$ bin.
The cross section and the kinematic variables
are frame invariant, and are measured in the Breit frame, defined to be
the frame where the photon energy vanishes.
In this frame the target fragmentation region ($x_p <0$) contains the proton remnants, while
the struck parton fragments into the current fragmentation region ($x_p > 0$), and the latter process
is equivalent to the fragmentation of a parton into an event hemisphere in $e^+ e^-$ reactions.
The factorization theorem dictates that, at leading twist,
the highly virtual photon undergoes hard scattering with a parton in the proton moving in the same direction and carrying away
an energy / momentum fraction $y$.
One of the partons produced in this scattering undergoes fragmentation to the observed hadron $h^\pm$ 
moving in the same direction and carrying away an energy / momentum fraction $z$.
In other words, after the change of integration variables $z\rightarrow x_p/z$ and $y\rightarrow x/y$,
the factorized cross section in the numerator of Eq.\ (\ref{XSmeas}) takes the form
\be
\begin{split}
\frac{d \sigma^{{\rm proton}\ h^\pm}}{dx_p dx dQ^2}(x,x_p,Q^2)
=&\int_{x_p}^1 \frac{dz}{z} \int_x^1 \frac{dy}{y} \sum_{ij} \frac{d \sigma^{ij}}{dz dy dQ^2}
\left(y,z,\frac{Q^2}{\mu^2},a_s(\mu^2)\right)\\
&\times
f_i^{\rm proton} \left(\frac{x}{y},\mu^2\right)
D_j^{h^\pm} \left(\frac{x_p}{z},\mu^2\right),
\label{xsinfacttheor}
\end{split}
\ee
where $f_i^{\rm proton}$ is the parton distribution function (PDF) of parton $i$ in the proton,
$D_j^{h^\pm}$ the FF of parton $j$ to $h^\pm$, $d\sigma ^{ij}$ the equivalent
factorized partonic observable given to NLO in Ref.\ \cite{Altarelli:1979kv}, $\mu$ the factorization / 
renormalization scale which distinguishes the soft from the hard subprocesses and
$a_s(\mu^2)=\alpha_s (\mu)/(2\pi)$.

\section{Comparisons with data}

\begin{figure}[h!]
\parbox{.49\linewidth}{
\begin{center}
\includegraphics[width=7.2cm]{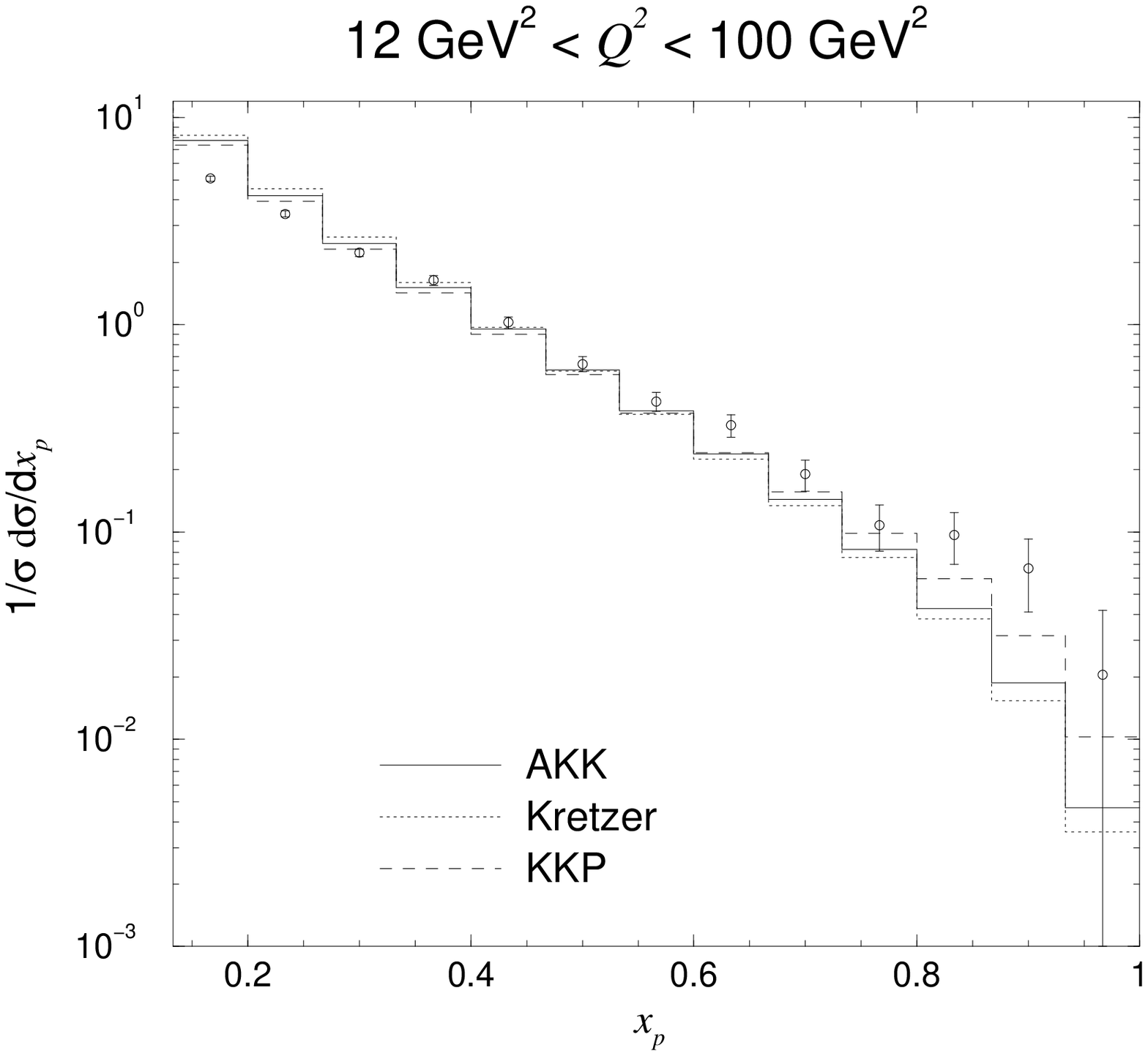}
\end{center}
}\hspace{0cm}
\parbox{.49\linewidth}{
\begin{center}
\includegraphics[width=7.2cm]{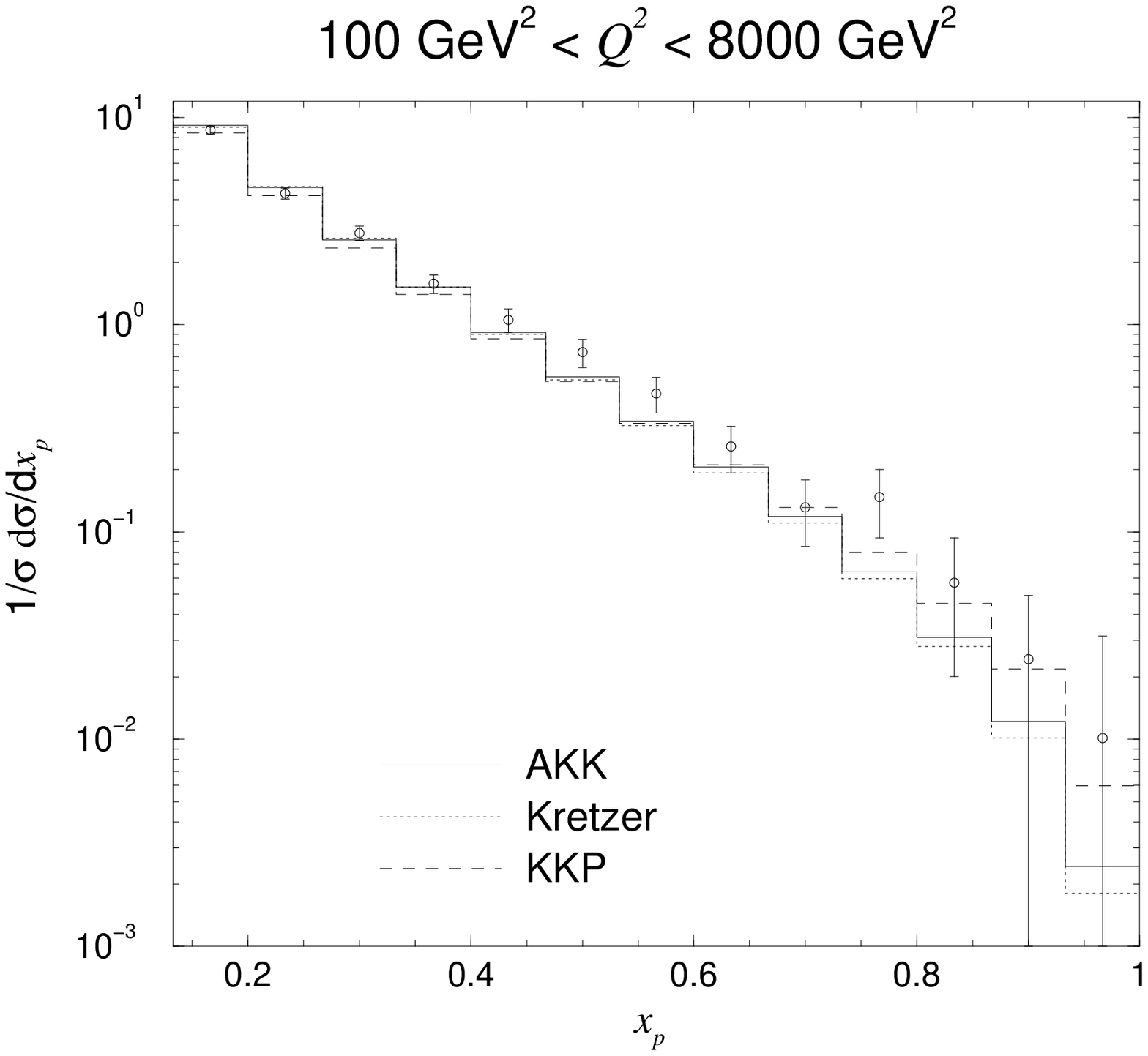}
\end{center}
}
\caption{Comparisons of theoretical predictions using the AKK, Kretzer
and KKP FF sets with the $x_p$ distributions from H1 \cite{Adloff:1997fr}.}
\label{H1old}
\end{figure}

\begin{figure}[h!]
\parbox{.49\linewidth}{
\begin{center}
\includegraphics[width=7.2cm]{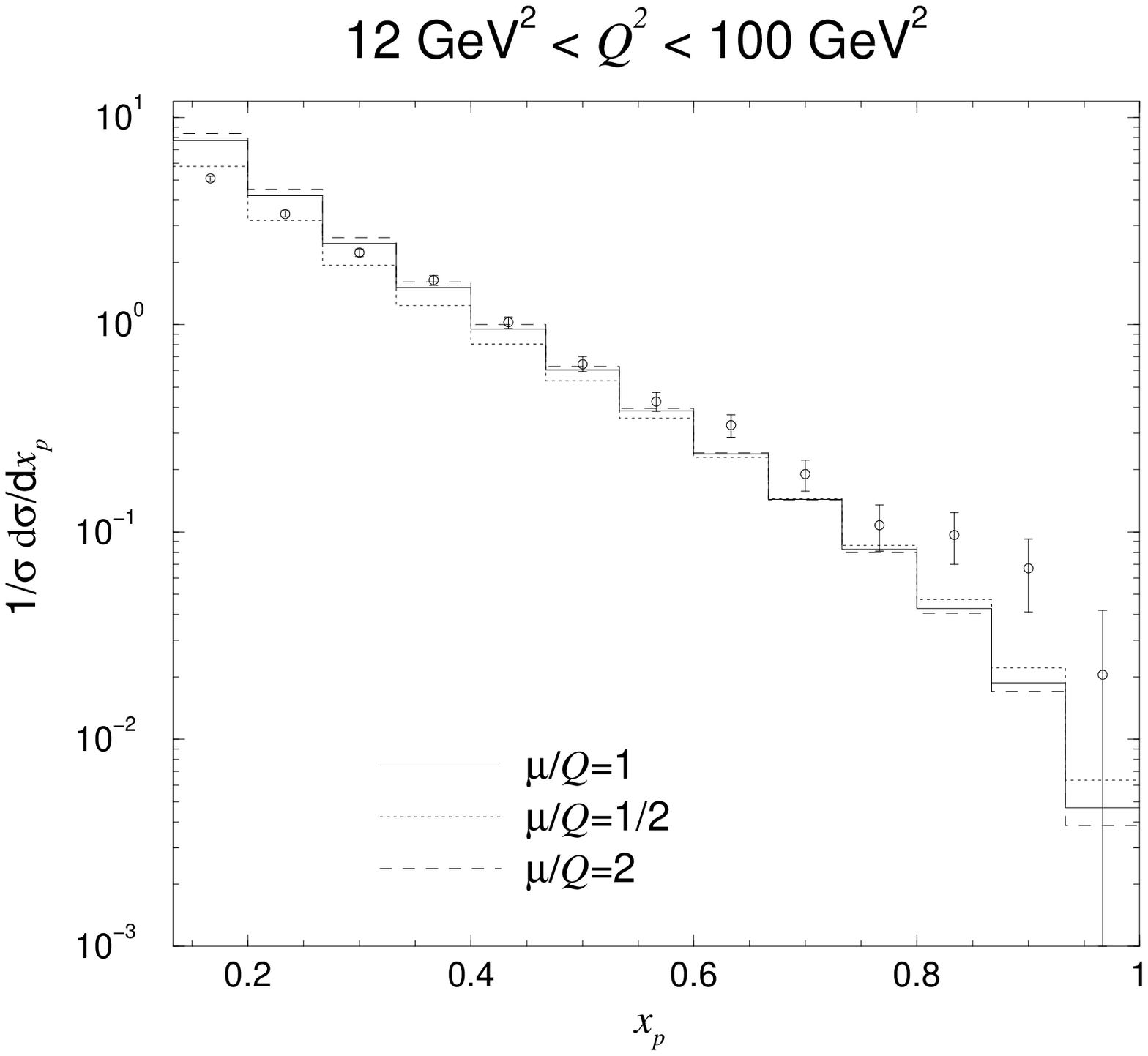}
\end{center}
}\hspace{0cm}
\parbox{.49\linewidth}{
\begin{center}
\includegraphics[width=7.2cm]{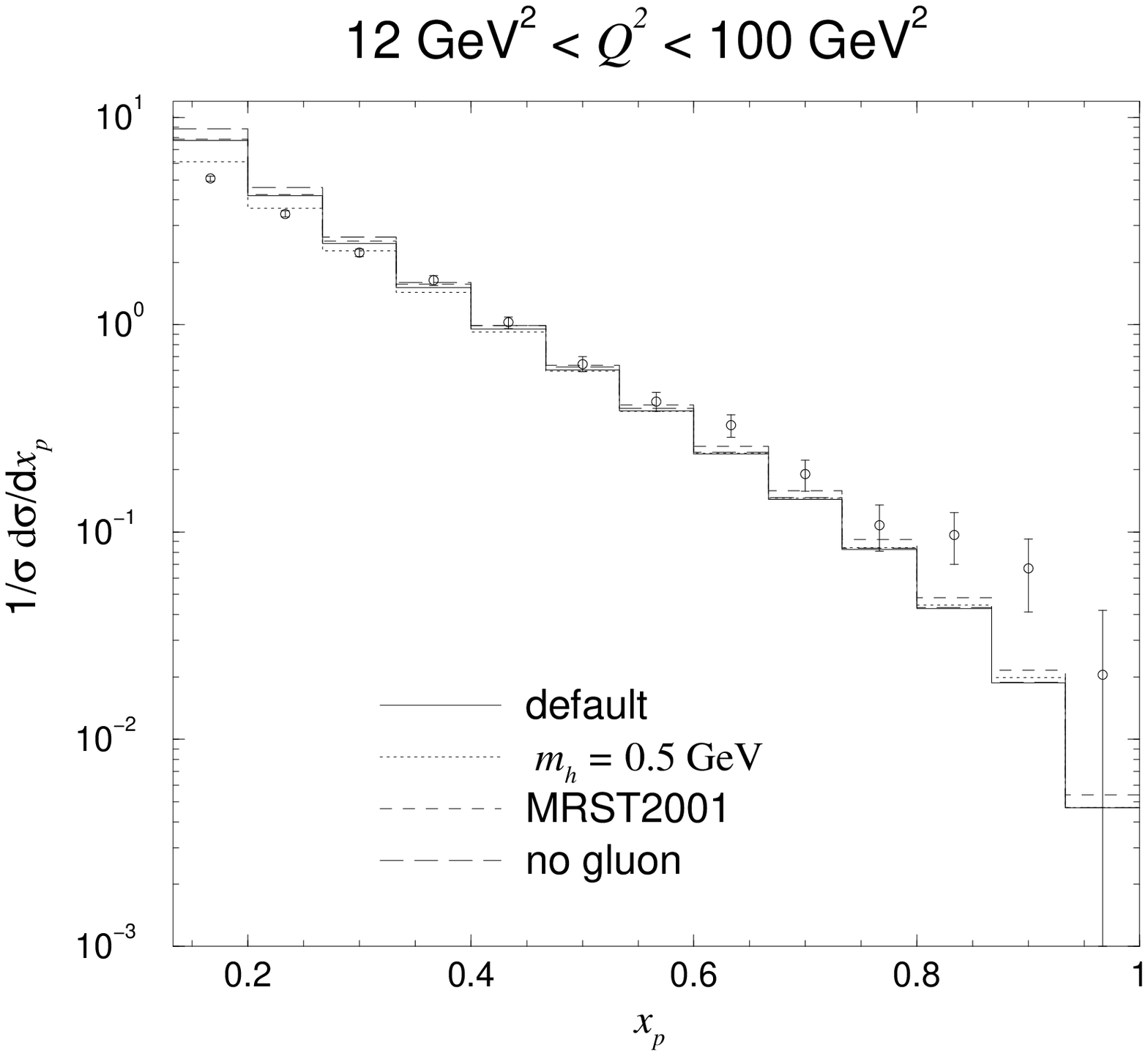}
\end{center}
}
\caption{As in Fig.\ \ref{H1old}, using only the AKK FF set.
Left: The modifications arising from scale variation.
Right: The modifications to the default predictions (solid line)
arising from the replacement of the CTEQ6M PDF set by the
MRST2001 PDF set of Ref.\ \cite{Martin:2001es}, from the removal
of the evolved gluon, and from the incorporation of the hadron mass effect are shown.}
\label{H1old_combine}
\end{figure}

At leading order in $a_s$, eq.\ (\ref{XSmeas}) becomes
\be
F^{{\rm proton}\ h^\pm}({\rm cuts},x_{pA},x_{pB})=\frac{\int_{x_{pA}}^{x_{pB}} dx_p \sum_I e_{q_I}^2(Q^2) G_I(Q^2)
x_p D_I^{h^\pm} (x_p,Q^2)}{\sum_J e_{q_J}^2(Q^2) G_J(Q^2)},
\label{simpformofF}
\ee
where the parton labels $I,J$ are restricted to (anti)quarks $q_I$ only, which have electric charges $e_{q_I}$, 
and $G_I(Q^2)=\int_{\rm cuts} dx\ x f_I^{\rm proton}(x,Q^2)$.
In the limit that the $G_I$ become independent of $I$, the numerator of Eq.\ (\ref{simpformofF}) is equal to
the equivalent LO result for $e^+ e^- \rightarrow h^\pm+X$, and therefore
the two types of observables are distinguished only by the $G_I$.
If this discrepancy is small, a good description of HERA data
is expected using FFs obtained from fits to $e^+ e^-$ data, 
such as the AKK \cite{Albino:2005me}, Kretzer \cite{Kretzer:2000yf} and KKP \cite{Kniehl:2000fe} FF sets
\footnote{Since this work was completed, 3 further sets \cite{Hirai:2007cx} have been extracted using
improved theoretical and experimental input.}
if universality and fixed order (FO) perturbation theory are reliable.
Calculations using these sets for the H1 data are shown in Fig.\ \ref{H1old},
using the CTEQ6M PDF set \cite{Pumplin:2002vw} and the CYCLOPS program \cite{Graudenz:1996yg} here and throughout this work,
unless otherwise stated.
The strong disagreement between the FF sets at large $x_p$ most likely arises from large experimental
errors on the FFs due to poor constraints from  $e^+ e^-$ reaction data at large momentum fraction.
At high $Q^2$, the calculation for all 3 FF sets agrees well with the data.
Therefore, the disagreements at large and small $x_p$ values found with the lower $Q^2$ data 
may be due to effects beyond the FO approach at leading twist.
For example, resummation of soft gluon emission logarithms
that become large at small and large $x_p$ may be necessary to improve the calculation here.
This is illustrated in Fig.\ \ref{H1old_combine} (left) by the effect of scale variation on the calculation, being largest
at small and large $x_p$.
The effect of the observed hadron's mass $m_h$ is also important at small $x_p$ for low $Q^2$ values.
For non-zero hadron mass, one has to distinguish between momentum, energy, light cone momentum etc.,
which are all equal when the hadron mass is negligible.
According to the factorization theorem, the ``momentum'' fraction $z$ appearing in eq.\ (\ref{xsinfacttheor}) is 
the fraction of light cone momentum carried away from the fragmenting parton by the observed hadron in a frame 
in which the spatial momenta of the virtual photon and the detected hadron are parallel,
and $x_p=\xi_p (1-m_h^2 /(Q^2 \xi_p^2))$ 
should be replaced by the ratio of the hadron's to the virtual photon's light cone momentum, $\xi_p$.
Using this approach \cite{Albino:2006wz}, one finds that 
the experimentally measured quantity $d \sigma^{{\rm proton}\ h^\pm}/ dx_p dx dQ^2$ 
is related to the calculated quantity $d \sigma^{{\rm proton}\ h^\pm}/ d\xi_p dx dQ^2$ by
\be
\frac{d \sigma^{{\rm proton}\ h^\pm}}{dx_p dx dQ^2}(x,x_p,Q^2)=\frac{1}
{1+\frac{m_h^2 }{Q^2 \xi_p^2(x_p)}}\ \frac{d\sigma^{{\rm proton}\ h^\pm}}{d\xi_p dx dQ^2}(x,\xi_p(x_p),Q^2),
\ee
which shows clearly that hadron mass effects become important at small $x_p$ and low $Q^2$.
According to Fig.\ \ref{H1old_combine}, this correction improves the description in this region, if we compare the results of 
this figure with the low $Q^2$ results of Fig.\ \ref{H1old}.
The choice $m_h=0.5$ GeV represents an ``average'' mass for the light charged hadrons.
We do not incorporate mass effects for the proton of the initial state, since
this effect is expected to partially cancel between the numerator and denominator of eq.\ (\ref{XSmeas}).
By redoing the calculation with the MRST2001 PDF set \cite{Martin:2001es}, we see that the dependence on the choice of PDF set
is small, particularly at small $x_p$, most likely because these quantities are well constrained 
but also because any variations in them are partially canceled between the numerator and denominator of eq.\ (\ref{XSmeas}).
As for $e^+ e^-$ reactions, the dependence on the gluon FF is small, particularly at large $x_p$.

\begin{figure*}[h!]
\begin{center}
\includegraphics[width=13.5cm]{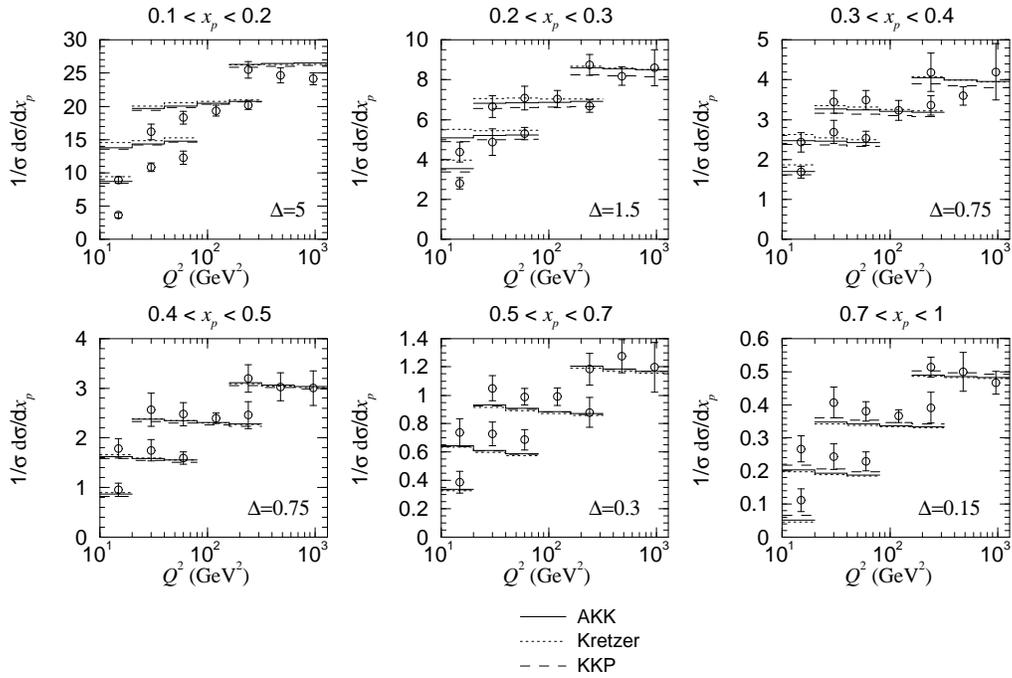}
\caption{As in Fig.\ \ref{H1old}, but for the ZEUS data \cite{Breitweg:1997ra}. 
Each data set is measured in a specific $x_p$-bin and, together with its predictions, is
shifted upward relative to the one below by the indicated value for $\Delta$.}
\label{ZEUSold2}
\end{center}
\end{figure*}

\begin{figure*}[h!]
\begin{center}
\includegraphics[width=13.5cm]{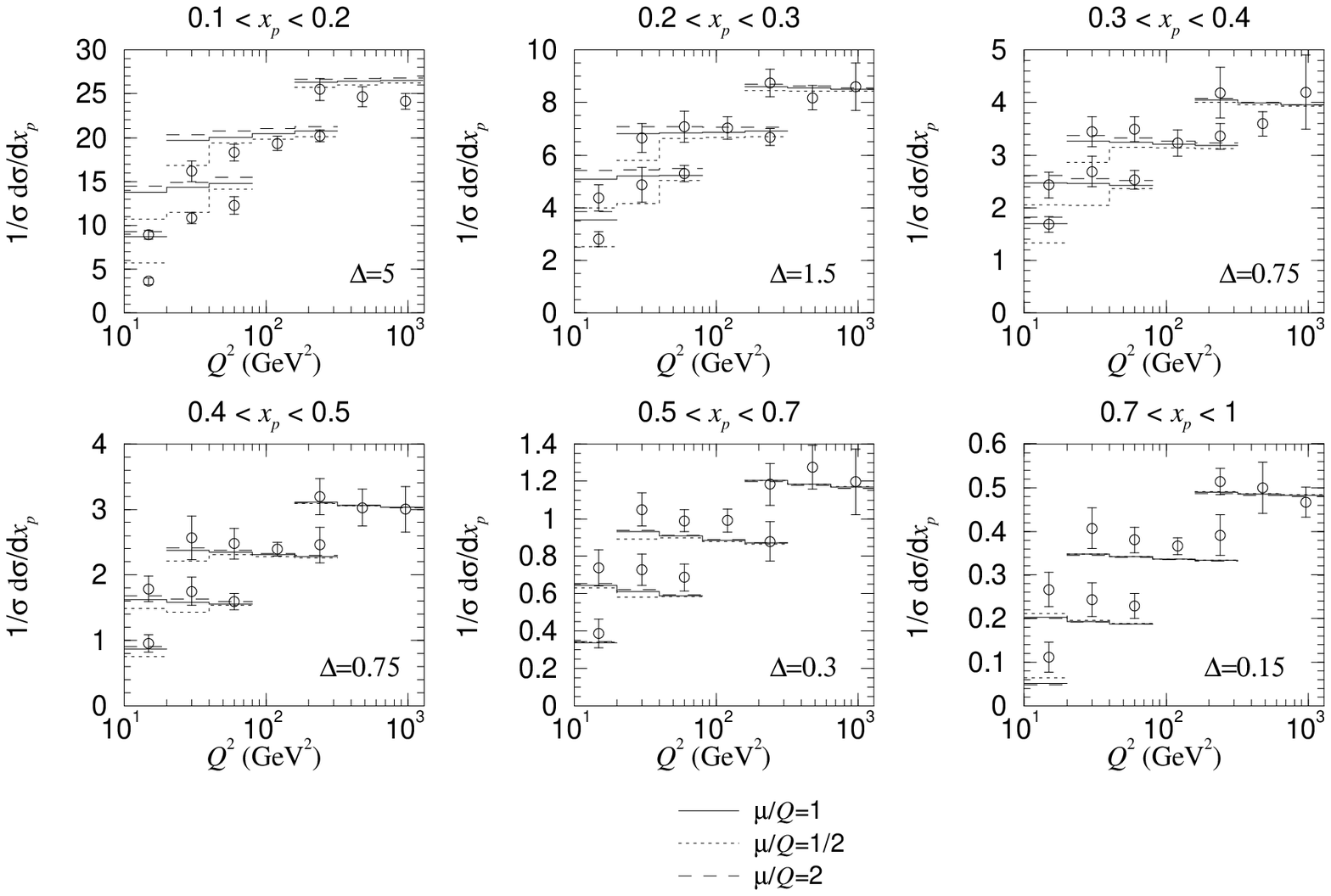}
\end{center}
\begin{center}
\includegraphics[width=13.5cm]{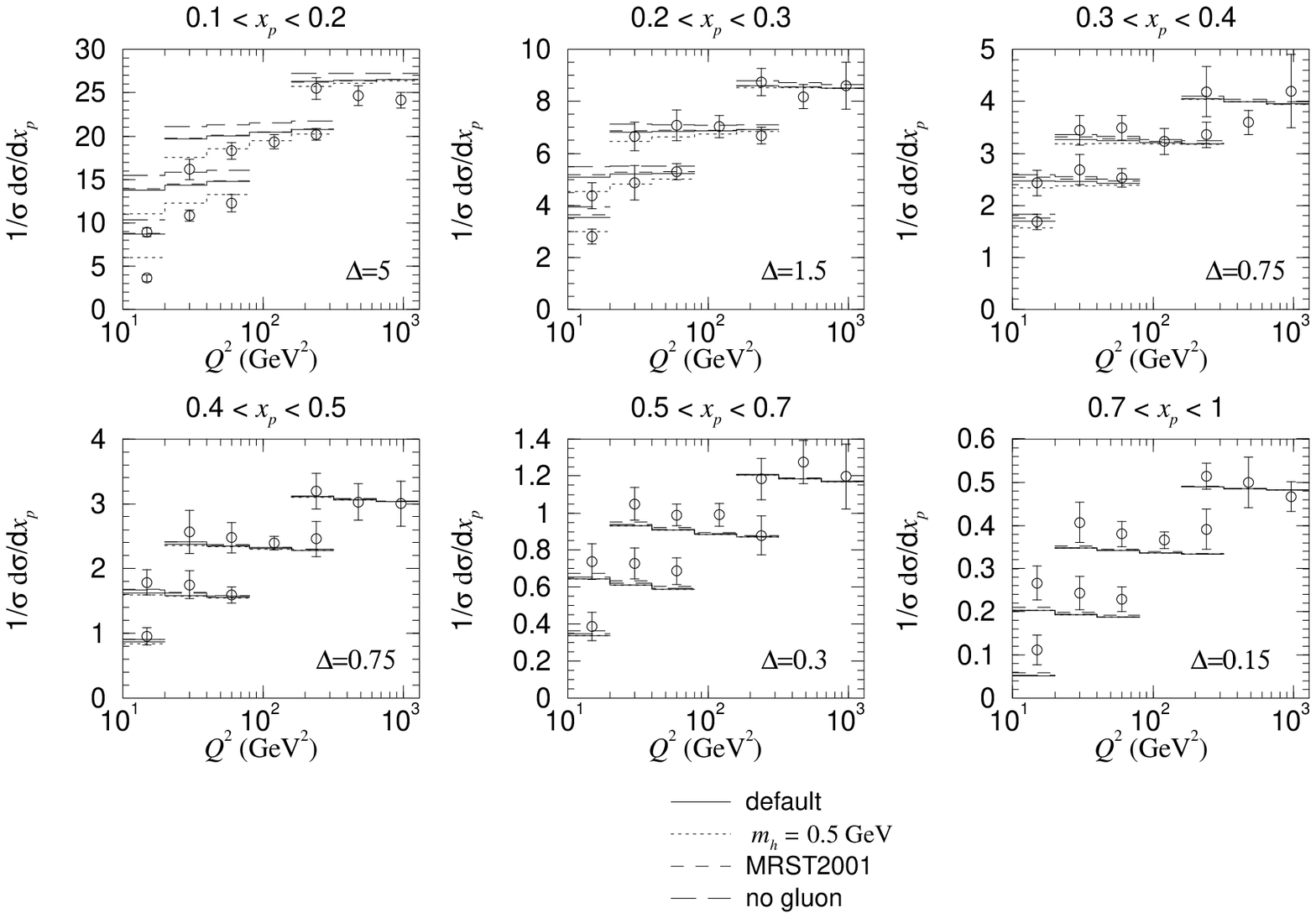}
\end{center}
\caption{As in Fig.\ \ref{H1old_combine}, but for the ZEUS data \cite{Breitweg:1997ra}.}
\label{ZEUSold2_combine}
\end{figure*}

To further verify these observations and inferences,
we perform similar calculations for the ZEUS data.
The different FF sets lead to similar results and good agreement with the data at large $Q^2$ 
and intermediate $x_p$ (Fig.\ \ref{ZEUSold2}).
The scale variation (Fig.\ \ref{ZEUSold2_combine}, top) generally decreases with increasing $Q^2$, 
and is largest for small $x_p$.
Both hadron mass effects and gluon fragmentation are most important 
at low $Q^2$ and small $x_p$ (Fig.\ \ref{ZEUSold2_combine}, bottom).

\section{Conclusions}

High $Q^2$ measurements of inclusive single hadron production at HERA are well described by perturbative QCD 
in the framework of the factorization theorem using available FF sets.
Although some disagreement is found with data at lower $Q^2$, we note that there is significant room for 
improvement in the theory in this region, such as hadron mass effects studied in this work, but also
resummation of the FO series at small and large $x_p$, higher twist effects
and quark mass effects.
Whether such effects are in fact relevant can be better verified by the effect of incorporating such data into global fits of FFs. 
More importantly, such data may also provide valuable information on the FFs' quark flavour components not constrained by 
$e^+ e^-$ reaction data, particularly since these type of HERA measurements may now be made very precisely \cite{Aaron:2007ds}.
However, in order to constrain FFs for each hadron species individually, and to ensure that the data is not contaminated
by light charged particles other than $\pi^\pm$, $K^\pm$ and $p/\overline{p}$, 
the hadron species of the HERA data need to be identified.

\section*{Acknowledgments}

The author thanks Bernd A.\ Kniehl, Gustav Kramer and Carlos Sandoval for collaboration.

%------------------------------------------------------------------------------
%       Bibliography
%------------------------------------------------------------------------------
%\bibliographystyle{heralhc} 
%{\raggedright

%}
\end{document}